\documentclass[newstyle,twocolumn,proceedings]{rmaa}

\usepackage{rmaacite}

\renewcommand{\P}[1]{%
\ifnum#1=1\hbox{OW~168--326E}\fi
\ifnum#1=2\hbox{OW~167--317}\fi
\ifnum#1=3\hbox{OW~163--317}\fi
\ifnum#1=5\hbox{OW~158--323}\fi
\ifnum#1=0\hbox{OW~171--334}\fi}
\makeatletter
\title{Cosmology with GTC.  
A combined mm-optical galaxy cluster survey}
\author{J. M. Diego\altaffilmark{1}, E. Mart\'\i nez-Gonz\'alez\altaffilmark{2}, 
       J.L. Sanz\altaffilmark{2}, N. Benitez\altaffilmark{3}, J. Silk\altaffilmark{1}
   \altaffiltext{1}{University of Oxford}
   \altaffiltext{2}{Instituto de F\'\i sica de Cantabria.}
   \altaffiltext{3}{The Johns Hopkins University.}}

\fulladdresses{
\item Dept. Astrophysics. University of Oxford. Keble road 1, OX1 3RH Oxofrd, UK 
      (jmdr@astro.ox.ac.uk).
\item Instituto de F\'\i sica de Cantabria. Avda. Los Castros. 39005 Santander, Spain.
\item Dept. of Physics \& Astronomy. The Johns Hopkins University. 
      3400 North Charles Street. Baltimore, MD 21218-2686. US.
}

\shortauthor{Diego et al.}
\shorttitle{Cosmology with GTC}


\abstract{%
   The CMB Planck satellite (ESA) will provide a catalogue of $\approx 30000$ 
   galaxy clusters (Sunyaev-Zel'dovich effect) between redshift 0 
   and $\approx$ 2-3 (depending on the cosmological model). 
   This hughe catalogue of clusters will allow detailed 
   studies of the evolution of the cluster population with redshift which will 
   constrain the cosmological model. The identification of the high redshift clusters 
   should be done with a large telescope (like GTC) since Planck data alone will not 
   provide any clue about the redshift of the clusters. In this work we show that  
   an optical follow up of only $\approx$ 300 clusters 
   (randomly selected from the Planck catalogue) is needed in order to distingish 
    models with or without a cosmological constant.} 

\resumen{El satelite para el estudio del CMB, Planck (ESA), originara un catalogo 
    con  $\approx 30000$ c\'umulos de galaxias (efecto Sunyaev-Zel'dovich) entre redshifts 
    0 y  $\approx$ 2-3 (dependiendo del modelo cosmol\'ogico). Tal extenso cat\'alogo 
    de cumulos permitira llevar a cabo estudios muy detallados de la evoluci\'on de 
    la poblacion de c\'umulos con el redshift lo cual permitir\'a a su vez afinar 
    en la estimaci\'on del modelo cosmol\'ogico. No obstante, la identificaci\'on de 
    los c\'umulos mas distantes tendr\'a que ser llevada a cabo con telescopios 
    opticos de gran diametro (como GTC) ya que Planck no sera capaz de dar una 
    estimaci\'on del redshift de los c\'umulos. En este trabajo mostramos como s\'olo 
    hace falta identificar una peque\~{n}a porcion de  $\approx$ 300 c\'umulos (que 
    hayan sido seleccionados aleatoriamente del cat\'alogo de Planck) para distinguir 
    modelos que tienen o no tienen constante cosmol\'ogica.  }

\listofauthors{J.~M.~Diego, E.~Mart\'\i nez-Gonz\'alez, N.~Benitez, J.~L.~Sanz, J.~Silk.}
\indexauthor{Diego, J.~M}
\indexauthor{Mart\'\i nez-Gonz\'alez, E.}
\indexauthor{Benitez, N.}
\indexauthor{Sanz, J.~L.}
\indexauthor{Silk, J.}

\begin{document}


\maketitle


\section{Introduction}
\label{sec:intro}
Clusters of galaxies have been widely used as cosmological probes. 
Their modeling can be easily understood as they are the final stage of 
the linearly evolved primordial density fluctuations. 
As a consequence, it is possible to describe, as a function of the 
cosmological model, the distribution of clusters and their evolution, 
the {\it mass function}, which is usually used as a cosmological test .
Therefore, a detailed study of the cluster mass function will provide 
very useful information about the underlying cosmology.\\
Unfortunately, cluster masses can not very well determined for intermediate-high 
redshift clusters and even for low redshift ones the error bars are 
still significant. 
However, instead of the mass function, it is possible to study the cluster population 
through other functions like the X-ray flux or luminosity functions, the 
temperature function or the Sunyaev-Zel'dovich effect (SZE hereafter) function. 
The advantage of these functions compared with the mass
function is that, in these cases, the estimation of the X-ray fluxes, 
luminosities, temperatures or SZE decrements of the clusters is less affected by 
systematics than the mass estimation. 
The largest catalogue of clusters in the next years will be provided by the CMB 
satellite Planck. It is expected that Planck data will contain about 30000 
detectable clusters (see Diego et al. 2002). \\
The decrement in the CMB temperature due to the SZE  
is independent of redshift. Therefore the most distant clusters could be detected 
through the SZE. Hence, the SZE is the perfect way to look at those high 
redshift clusters. It is in the high redshift interval where the differences among 
the cosmological models are more evident when one looks at the cluster population. \\
Unfortunately, through the SZE it is not possible to measure the redshift of the 
clusters and an independent observation (in the optical waveband for instance) of the 
clusters is needed in order to estimate their redshifts. \\
Since many clusters will be at high redshift, a telescope with a large diameter 
(like GTC) will be needed in order to identify those high-z clusters. 
However, due to the large size of the Planck catalogue, any proposal which 
attempts to make use of a 10-m class telescope to identify 30000 clusters would be 
rejected (even in the case the proposal plans to observe only the high-z 
clusters there would be thousands of them !). 
In the next sections we will show how an SZE-selected optical survey made with GTC 
of only a small portion of the clusters in the Planck catalogue could be enough 
to obtain important cosmological constraints.

\section{A combined mm-optical survey} 
\label{sec:survey}

At the end of this decade, the Planck satellite will carry out a full-sky survey 
in order to measure with a high sensitivity the CMB. 
However, other emissions (not only due to the CMB) will also be  
present in the data. Most of this non-CMB emissions will come from our own 
Galaxy (dust, free-free, synchrotron) but there will be also some extra-galactic 
emissions (point sources and the SZE). The wide frequency range, angular 
resolution and high sensitivity of Planck will allow to detect $\approx$ 30000 
clusters. \\
Observing galaxy clusters in the mm sub-mm band through the SZE has a unique 
advantage. The amplitude of the decrement in the center of the cluster is independent 
of its redshift. Due to this fact, the catalogue of clusters obtained 
by Planck will have a privileged selection function and the proportion 
of high-to-low redshift clusters will be maximum if we compare the catalogue 
with others obtained in optical or X-ray surveys. \\
Galaxy clusters are the best tracers of the large scale structure and by studying the 
evolution of their abundance with redshift it is possible to impose very strong 
constraints on the cosmological model. 
\begin{figure}
  \includegraphics[width=\columnwidth]{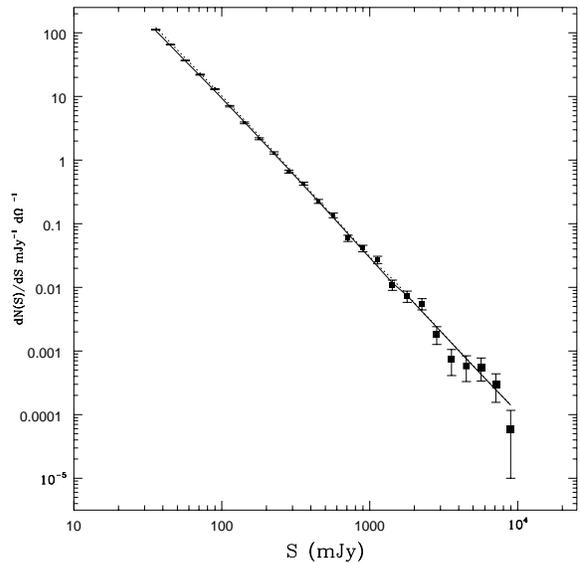}
  \caption{Cluster number counts for Planck. Two models are plotted for comparison. 
           OCDM model is the solid line ($\Omega = 0.3$, $\Lambda = 0.0$) and  
           $\Lambda$CDM the dotted line ($\Omega = 0.3$, $\Lambda = 0.7$). The 
           data is consistent with both models. Only the number counts as a function of 
           redshift can distinguish both models.}
  \label{fig:Fig1}
\end{figure}
For instance, using the local abundance of clusters it has been possible 
to find a strong correlation between the amplitude of the power spectrum ($\sigma _8$) 
and the matter density ($\Omega$). However, all the models in that correlation having 
different values of $\sigma _8$ and $\Omega$ predict the same observed local abundance of 
clusters (within the error bars). Therefore, by using just the local abundance of 
clusters it is not possible for instance to rule out models with high or low 
values of $\Omega$. However this can be done if we go back on time and study 
the evolution of the cluster abundance with redshift. In this case, the further we 
go in redshift, the better are the constraints in the cosmological parameters. \\
\begin{figure}
  \includegraphics[width=\columnwidth]{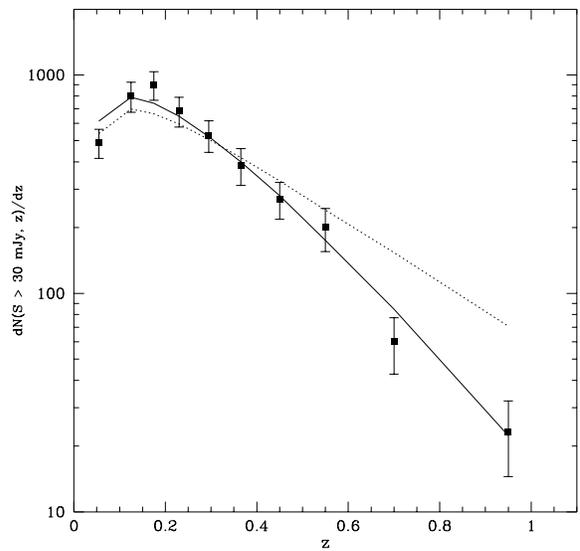}
  \caption{Evolution of the number counts for the subsample of 300 clusters  
           The data points were obtained from a Montecarlo simulation of 
           the $\Lambda$CDM model. The solid 
           line is the mean expected number counts for the same model ($\Lambda$CDM) 
           and the dotted line is the corresponding expected number counts for the 
           OCDM model. The OCDM model is excluded at 3 $\sigma$ level. }
  \label{fig:Fig2}
\end{figure}
As we have mentioned, the selection function of Planck will be privileged in the sense 
that the final catalogue will have a large proportion of high redshift clusters.
It is interesting to exploit this fact and study the cosmological implications 
of such a large and redshift-independent catalogue. 
Unfortunately, since the SZE is independent of redshift, it will not be possible to 
determine the redshift of the clusters by just looking at their SZE emission. 
This fact will limit the kind of cosmological studies which can be done with the 
Planck catalogue since the only observable will be the flux of the cluster. 
In Fig. \ref{fig:Fig1} we show the expected number counts for the Planck catalogue  
as a function of the observed flux.\\
Despite the large number of clusters in the catalogue, 
this number will not be enough to discriminate between a model with a cosmological 
constant and a model without cosmological constant. \\
However, both models can be distinguished if one uses the evolution of 
the number counts with redshift. To build this data one needs, obviously, to estimate 
first the redshift of the clusters and since this can not be done through the SZE 
one should make an independent optical observation for each one of the clusters. 
This can be a huge task if one attempts to measure the redshift for each one 
of the expected 30000 clusters. Some of them will be nearby clusters and they can be 
easily identified in previous surveys like the Sloan. Others at intermediate 
redshift could be observed with medium-size telescopes. There will be however, a large 
number of high-z clusters. In these cases, a large telescope like GTC will be needed. 
But even observing only the high-z clusters, their expected number is still too large 
to make a project like this one possible. 
However, instead of trying to identify all the Planck cluster catalogue, one could try 
to estimate the redshifts for only a small portion of it and build the number counts 
(as a function of z) from them. The question now is, how small should be the subsample 
of clusters if we want to extract useful information about the cosmological 
information form that subsample ? \\
In Diego et al. (2001) we computed such number and we found that due to the particular 
selection function of Planck, by randomly selecting a subsample of only 300 clusters 
we could discriminate between $\Lambda$CDM and a OCDM models. This is an important 
conclusion since it tells us that we do not need to identify all the clusters in the 
Planck catalogue. The selection function of Planck is such that, a random subsample of 300 
clusters contain enough high-z clusters to make possible the distinction between 
the two models. \\
As we suggested before, part of those 300 clusters could be identified with 
existing cluster catalogues. Others could be identified using medium-size telescopes 
and only a small portion of them ($\approx 20 \%$) would require a telescope like 
GTC. In the same paper we calculated that, in order to estimate the photometric 
redshifts for the most distant clusters with GTC, we should observe them 
with two hours of integration time (8 bands and 900s per band) per cluster. 
These numbers show that this is a feasible project and the evolution of the 
number counts of the SZE-selected subsample of 300 clusters could be determined. \\
In Fig. \ref{fig:Fig2} we show the expected number counts as a function of redshift 
for a Montecarlo realization of the $\Lambda$CDM model. Also plotted are the mean 
number counts for the $\Lambda$CDM (solid line) and for the OCDM model (dotted line).
As we mentioned earlier, with just 300 clusters the OCDM model could be excluded 
by the high-z bins.  \\
\begin{figure}
  \includegraphics[width=\columnwidth]{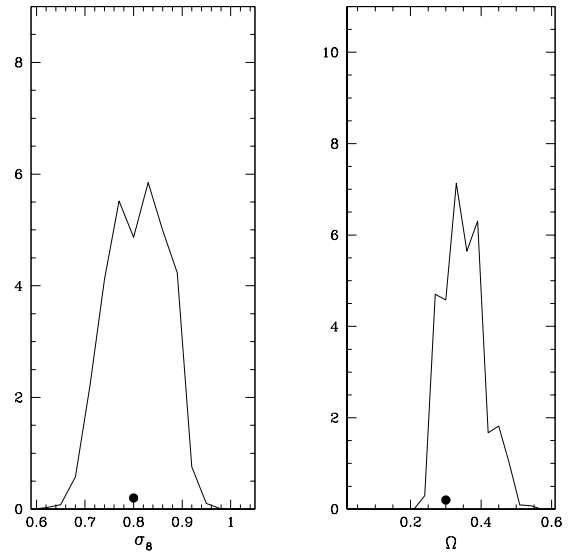}
  \caption{Cosmological constraints (marginalized probability) 
   obtained after combining the Planck number 
   counts (Fig. 1) with the evolution of the number counts of the optically 
   observed subsample of 300 clusters (Fig. 2). The fiducial model used to simulate 
   the two data sets is indicated by the big black dots. }
  \label{fig:Fig3}
\end{figure}
By combining the two data sets (number counts of Planck (Fig 1) and evolution of 
the number counts of the subsample (Fig. 2)) it is possible to constraint the 
cosmological parameters of the $\Lambda$CDM model. \\
The first data set is very good to constrain the $\sigma _8 - \Omega$ correlation. 
The second data can be used to break the previous degeneracy between both parameters.
The result of combining the Planck number counts (30000 clusters, Fig. 1) 
with the GTC followup (300 clusters, Fig. 2) can be seen in Fig. \ref{fig:Fig3}. \\

\section{conclusions}
We have seen how combining two very different instruments (Planck and GTC) it is 
possible to constraint the cosmological model. 
Planck will provide a unique way of selecting distant clusters. However a 
subsequent redshift estimation will be needed. We have seen that a subsample 
(randomly selected from the Planck catalogue) of 300 identified clusters  is 
large enough in order to distinguish between a $\Lambda$CDM and a OCDM model. 
Among these 300 clusters there will be many of them which can be identified 
using existing galaxy cluster catalogues or medium-size telescopes but there 
will be a portion ($\approx 20 \%$) which will require a telescope like GTC. 
A survey made with GTC over the SZE-selected subsample will allow to 
constrain the cosmological model in an independent test. 

\acknowledgements
This work has been supported by the 
Spanish DGESIC Project  
PB98-0531-C02-01, FEDER Project 1FD97-1769-C04-01, the 
EU project INTAS-OPEN-97-1192, and the RTN of the EU project   
HPRN-CT-2000-00124. \\
JMD acknowledges support from its Marie Curie Fellowship 
of the European Community programme {\it Improving the Human Research 
Potential and Socio-Economic knowledge} under 
contract number HPMF-CT-2000-00967.



\begin{thebibliography}

\bibitem[Diego<2001>]D Diego J.M., Mart\'\i nez-Gonz\'alez E., Sanz J.L., Benitez N., Silk J.,
         MNRAS accepted. astro-ph/0103512

\bibitem[Diego<2002>]D Diego J.M., Vielva P., Mart\'\i nez-Gonz\'alez E., Silk J., Sanz J.L., 
         MNRAS submitted. astro-ph/0110587.

\end{thebibliography}
\end{document}